\documentstyle[12pt,procsla,epsfig]{article}

  \catcode`\@=11
  \long\def\@makefntext#1{
  \protect\noindent \hbox to 3.2pt {\hskip-.9pt  
  $^{{\ninerm\@thefnmark}}$\hfil}#1\hfill}              
  
  \def\@makefnmark{\hbox to 0pt{$^{\@thefnmark}$\hss}}  
        
  \def\ps@myheadings{\let\@mkboth\@gobbletwo
  \def\@oddhead{\hbox{}
  \rightmark\hfil\ninerm\thepage}   
  \def\@oddfoot{}\def\@evenhead{\ninerm\thepage\hfil
  \leftmark\hbox{}}\def\@evenfoot{}
  \def\sectionmark##1{}\def\subsectionmark##1{}}
  
\begin{document}
  
  \centerline{\normalsize\bf ATMOSPHERIC NEUTRINOS AT SUPER-KAMIOKANDE}
  
  \centerline{\footnotesize KATE SCHOLBERG}
  \baselineskip=13pt
  \centerline{\footnotesize\it Boston University, Dept. of Physics, 590 Commonwealth Ave.}
  \baselineskip=12pt
  \centerline{\footnotesize\it Boston, MA 02215, USA}
  \centerline{\footnotesize E-mail: schol@budoe.bu.edu}
  \vspace*{0.3cm}
  \centerline{\footnotesize for}
  \vspace*{0.3cm}
  \centerline{\footnotesize THE SUPER-KAMIOKANDE COLLABORATION}
  
  \vfill \vspace*{0.9cm} \abstracts{ In 1998, the Super-Kamiokande
 announced evidence for the observation of neutrino oscillations based
 on measurements of the atmospheric neutrino flux\cite{nuosc}.  This
 paper presents the updated results for fully and partially-contained
 events with 736 days of data, as well as upward-going muon results and a
 global analysis.  Preliminary interpretations of the results in
 terms of various two-flavor oscillation hypotheses are presented.}
   
  \normalsize\baselineskip=15pt
  \setcounter{footnote}{0}
  \renewcommand{\thefootnote}{\alph{footnote}}
  \section{Introduction}

  \subsection{Atmospheric Neutrinos}
  
  Atmospheric neutrinos are produced by collisions of cosmic rays with
  atoms in the Earth's upper atmosphere.  These collisions produce
  showers which include neutrinos with energies in the 0.1~GeV to
  100~GeV range.  Although calculations of the absolute flux of atmospheric
  neutrinos~\cite{honda,bartol} have fairly large uncertainties
  ($\sim20$\%), the ratio of muon to electron flavor content is much
  better known (to $\sim5$\%), since it arises from the decay chain
  $\pi\rightarrow\mu\nu_{\mu},\mu\rightarrow e\nu_{\mu}~\nu_{e}$.  
  The muon to electron flavor ratio is
  expected to be about 2 (one $\nu_e$ from muon decay for every two
  $\nu_{\mu}$), gradually rising at energies above about 5~GeV as the
  muons from $\pi$ decay have enough energy to survive to the surface
  of the earth and range out before decaying.
  
Because the attenuation of even high energy neutrinos in the earth is so
small, neutrinos produced all around the atmosphere can be detected.
Their pathlengths range from about 15~km for neutrinos produced
directly overhead, to 13000~km for neutrinos produced directly below.
The pathlength of the neutrino can be inferred from the direction of
the Cherenkov cone of its produced lepton.  Therefore, by measuring
the zenith angle, energy and flavor of atmospheric neutrinos, we can
test the neutrino oscillation hypothesis, which gives the probability
of flavor conversion:

\begin{equation}
P_{f\rightarrow g} = \sin^2 2\theta \sin \left( \frac{1.27 \Delta m^2 ({\rm eV}^2) L ({\rm km})}{E_{\nu}({\rm GeV})}\right).
\end{equation}

  \subsection{The Super-Kamiokande Detector}

The Super-Kamiokande (Super-K) experiment is a large 
water Cherenkov detector located in Mozumi, Japan.  Its total mass of
ultra-pure water is 50~kton, divided into two concentric cylinders: an
inner volume with its inside surface covered by 11146
inward-looking 50~cm photomultiplier tubes, and an outer volume
serving as entering particle shield and veto with 1885 outward-looking
20~cm phototubes.  The fiducial mass of the inner volume (2~m away from
the walls) is 22.5~kton.

\section{Atmospheric Neutrinos with Super-K}

High energy neutrinos produced in the atmosphere are detected via
their interactions with nuclei in the water.  The easiest $\nu$ interaction
to tag and study in a water Cherenkov detector is the charged current (CC)
quasi-elastic interaction:
\begin{equation}
{\nu_{l}}~+~N~\rightarrow~l~+~N',
\end{equation}
where the flavor $l$ of the outgoing lepton tags the flavor of the
incoming neutrino.  The flavor of this lepton is determined using the
Cherenkov ring: events are identified as ``$\mu$-like''
(non-showering) or ``$e$-like'' (showering) based on a likelihood
analysis of the Cherenkov light around the cone.  At higher energies
(more than a few GeV), more complicated single and multi-pion products
from both charged current and neutral current (NC) neutrino
interactions with nuclei predominate.

Atmospheric neutrinos are divided into three classes for Super-K:

\begin{enumerate}
\item \textbf{Fully-contained events (FC):} events for which the
products of the neutrino interaction are completely contained within
the inner detector volume.  These are further subdivided into
``sub-GeV'' (with visible energy $<$1.33 GeV) and ``multi-GeV''
($>$1.33 GeV) events.  For the current analysis, single ring (mostly
quasi-elastic charged current) interactions are selected.  According
to Monte Carlo (MC), the FC $e$-like sample contains about
86\%~CC~$\nu_e$, 4\%~CC~$\nu_{\mu}$ and 10\% NC interactions; the FC
$\mu$-like sample contains about 96\%~CC~$\nu_{\mu}$, 4\%~CC~$\nu_{e}$
and $<$1\% NC interactions.

\item \textbf{Partially-contained events (PC):} events for which the
produced muon exits the inner detector volume (only muons are
penetrating enough). Such events comprise a 98\% pure sample of 
$\nu_{\mu}$ CC interactions, according to MC.  There is no single
ring selection for partially contained events.  The average energy of
a neutrino producing a PC event in Super-K is about
15~GeV.  

\item \textbf{Upward-going muons:}  upward-going muons result from neutrinos
which interact in the rock below the detector to produce a penetrating
muon.  As for PC events, we assume the neutrino
parent flavor is $\nu_{\mu}$, since only muon flavor neutrinos can
produce penetrating leptons.  These muons may either traverse the
entire detector or stop inside the volume.  Through-going upward muons
have average parent neutrino energies of about 100~GeV; stopping upward
muons have average parent neutrino energies of about 10~GeV.
(Of course, there are also downward-going neutrino-induced muons,
but we cannot distinguish these from the large background of 
entering downward-going cosmic ray muons).

\end{enumerate}

Details of the event selection and reconstruction (energy, vertex
position, direction, and number of rings), and of the MC
are described elsewhere~\cite{subgev,multigev,upmu,mmthesis}.

  \section{Recent Results}
  \subsection{What's New}

The new preliminary data sample, approved in December 1998, is described in
Table~\ref{tab:whatsnew}.

  \begin{table}[h]
	\centering
  \tcaption{Summary of June 1998 and December 1998 data and MC samples.}\label{tab:whatsnew}
  \vspace*{13pt}
  \small
  \begin{tabular}{||c|c|c||}\hline\hline
  \multicolumn{3}{||c||}{\textbf{DATA}} \\ \hline \hline
  &  June 1998 & December 1998\\ \hline
  \multicolumn{1}{||l|}{Analyzed ivetime:} & {} &{}\\ 
  \multicolumn{1}{||l|}{\hspace{0.5cm}Fully-contained} & 535 days & 736 days \\
  \multicolumn{1}{||l|}{\hspace{0.5cm}Partially-contained} & 535 days & 685 days \\
  \multicolumn{1}{||l|}{\hspace{0.5cm}Upward-going muons} & 516 days & 516 days \\

  \hline \hline
  \multicolumn{3}{||c||}{\textbf{MONTE CARLO}} \\ \hline \hline
   &  June 1998 & December 1998\\ \hline

   Generated livetime   & 10 years & 20 years \\
   High energy cross-sections & CCFR & GRV94 PDF\cite{grv} \\
   Solar epoch & solar cycle average & solar cycle minimum \\

  \hline \hline

  \end{tabular} 
  \end{table}

This new sample incorporates additional livetime for FC and PC events.
There were some minor improvements in reconstruction (fitting, ring
counting, etc.).  The MC, however, is a completely new sample.  The
most important differences are indicated in the table.  The net effect
of the change is an increase in the number and the fraction of
multi-ring events at higher energies.  Twenty years of livetime were
generated (previously there had been only 10 years).

\subsection{Fully and Partially-contained events}

  \subsubsection{East-West Effect}

The east-west effect results from the action of the geomagnetic field on
the charged cosmic ray primary particles.
Protons at the horizon with energies less
than about 50~GeV/c coming from the east are suppressed relative to
those coming from other directions, where the momentum cutoff is
smaller ($\sim$10~GeV/c).  To check the reliability of both our
experimental results and the atmospheric neutrino flux calculations,
we have selected a sample of neutrinos for which the east-west effect
is expected to be important: for which $\left|\cos\theta\right|<0.5$
and momentum is between 400 and 3000 MeV/c.

\begin{figure}
    \centering
    \mbox{\epsfig{figure=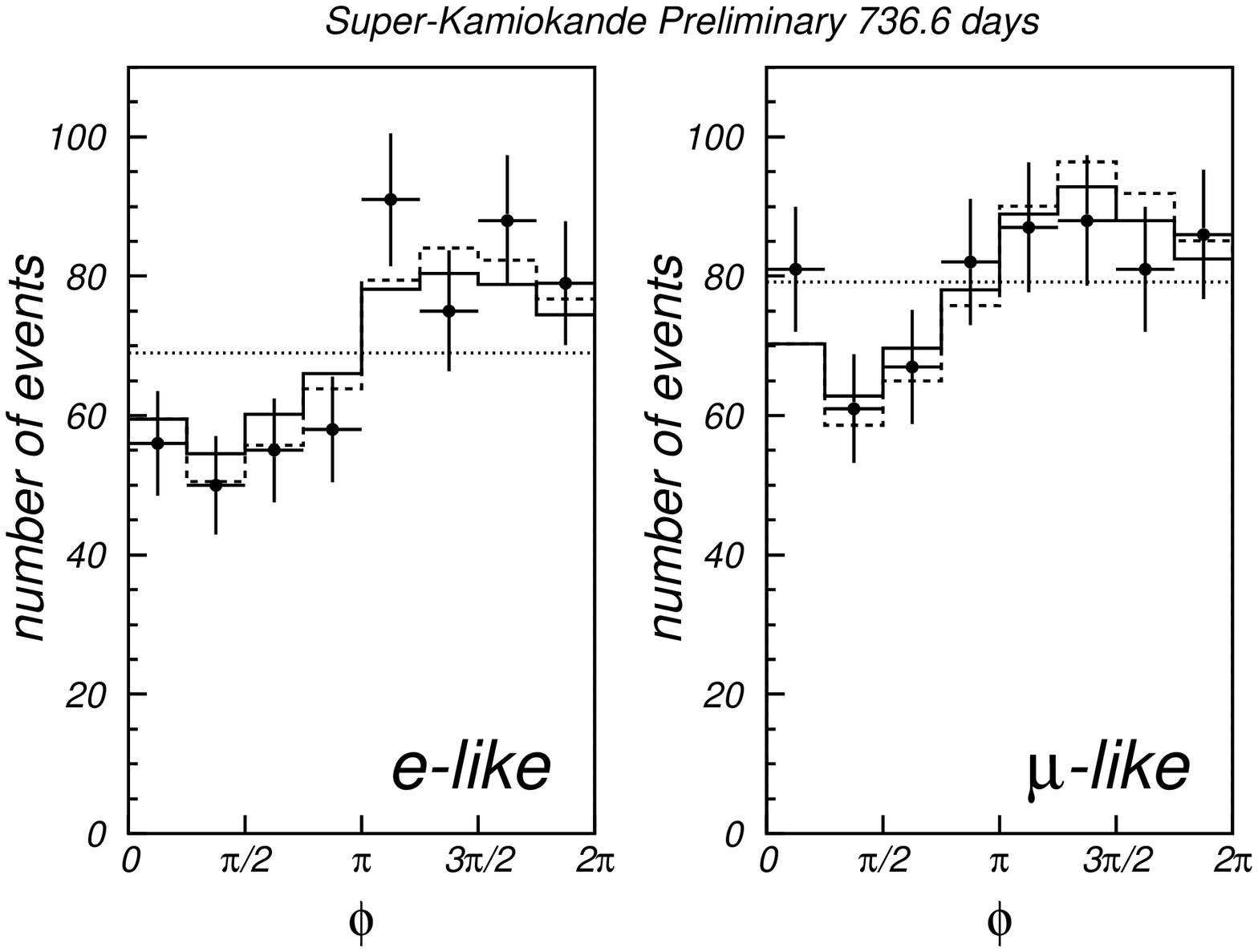,bb= 5 203 551 652,width=9.0cm}}
    \fcaption{The east-west effect: azimuthal distribution for
low-energy near-horizon $e$-like and $\mu$-like data, along with MC
predictions based on Honda (solid line) and Bartol (dashed line)
fluxes.  $\phi=0$ corresponds to particles going to the north and
$\phi=\pi$ corresponds to particles going to the south. The expected
deficit of neutrinos from the east due to the geomagnetic cutoff of
the charged primaries is observed. } \label{fig:eastwest}
\end{figure}

Figure~\ref{fig:eastwest} shows the azimuthal distribution for both
$e$-like and $\mu$-like events, along with two flux
predictions\cite{honda,geomag}.  The expected deficit from the east
is observed.  This analysis has been described in detail
elsewhere\cite{eastwest}.

  \subsubsection{Ratio of Ratios}

The measured flavor ratio is conventionally compared to expectation as
the ``ratio of ratios'', $R$, defined as
\begin{equation}
R~=~\frac{(N_{\mu}/N_e)_{DATA}}{(N_{\mu}/N_e)_{MC}},
\end{equation}
where $N_{\mu}$ and $N_{e}$ are numbers of $\mu$-like and $e$-like
events for data and MC.  These numbers (Honda flux\cite{honda}) are
given in Table~\ref{tab:bigR}, and the new values of $R$ are
\begin{equation}
R_{FC}~=~0.67~\pm~0.02(\rm{stat})~\pm0.05(\rm{sys})
\end{equation}
for sub-GeV FC events, and
\begin{equation}
R_{FC+PC}~=~0.66~\pm~0.04(\rm{stat})~\pm0.08(\rm{sys})
\end{equation}
for multi-GeV FC and PC events.  If there are no neutrino
oscillations, $R$ is expected to be 1.  However, $R$ is significantly
smaller than 1 for both sub-GeV and multi-GeV data.  A significantly
low value of $R$ has been measured by several experiments in the
past\cite{kam,imb}, and recently by the Soudan collaboration with an
iron-calorimeter detector\cite{soudan}.

\begin{table}[h]
\centering
  \tcaption{Numbers of events for FC and PC data and Monte Carlo.}
 \label{tab:bigR}
\vspace*{13pt}
\begin{minipage}[t]{6cm}
  \small
  \begin{tabular}{||c|c|c||}\hline\hline
  \multicolumn{3}{||c||}{\textbf{SUB-GEV}} \\ \hline \hline
  &  Data & MC\\ \hline
  \multicolumn{1}{||l|}{Single ring:} & {} &{}\\ 
  \multicolumn{1}{||l|}{\hspace{0.5cm}$e$-like} & 1607 & 1510.5 \\
  \multicolumn{1}{||l|}{\hspace{0.5cm}$\mu$-like} & 1617 & 2277.8\\
  \multicolumn{1}{||l|}{Multiple ring} & 1271 & 1614.3 \\ \hline
  \multicolumn{1}{||l|}{Total} & 4495 & 5402.5 \\
  \hline \hline
  \end{tabular} 
  \end{minipage}
  \begin{minipage}[t]{6cm}
  \small
  \begin{tabular}{||c|c|c||}\hline\hline
  \multicolumn{3}{||c||}{\textbf{MULTI-GEV}} \\ \hline \hline
  &  Data & MC\\ \hline
  \multicolumn{1}{||l|}{Single ring:} & {} &{}\\ 
  \multicolumn{1}{||l|}{\hspace{0.5cm}$e$-like} & 386 & 357.4 \\
  \multicolumn{1}{||l|}{\hspace{0.5cm}$\mu$-like} & 301 & 415.9\\
  \multicolumn{1}{||l|}{Multiple ring} & 737 & 
                                                  925.7 \\ \hline

  \multicolumn{1}{||l|}{Total} & 1424 & 1698.9 \\ \hline
  \multicolumn{1}{||l|}{PC ($\mu$-like)} & 374 & 
                                                  528.7 \\ \hline
  \hline 
  \end{tabular} 
  \end{minipage}
 \end{table}

  \subsubsection{Angular Distributions}

\begin{figure}
    \centering
\mbox{\epsfig{figure=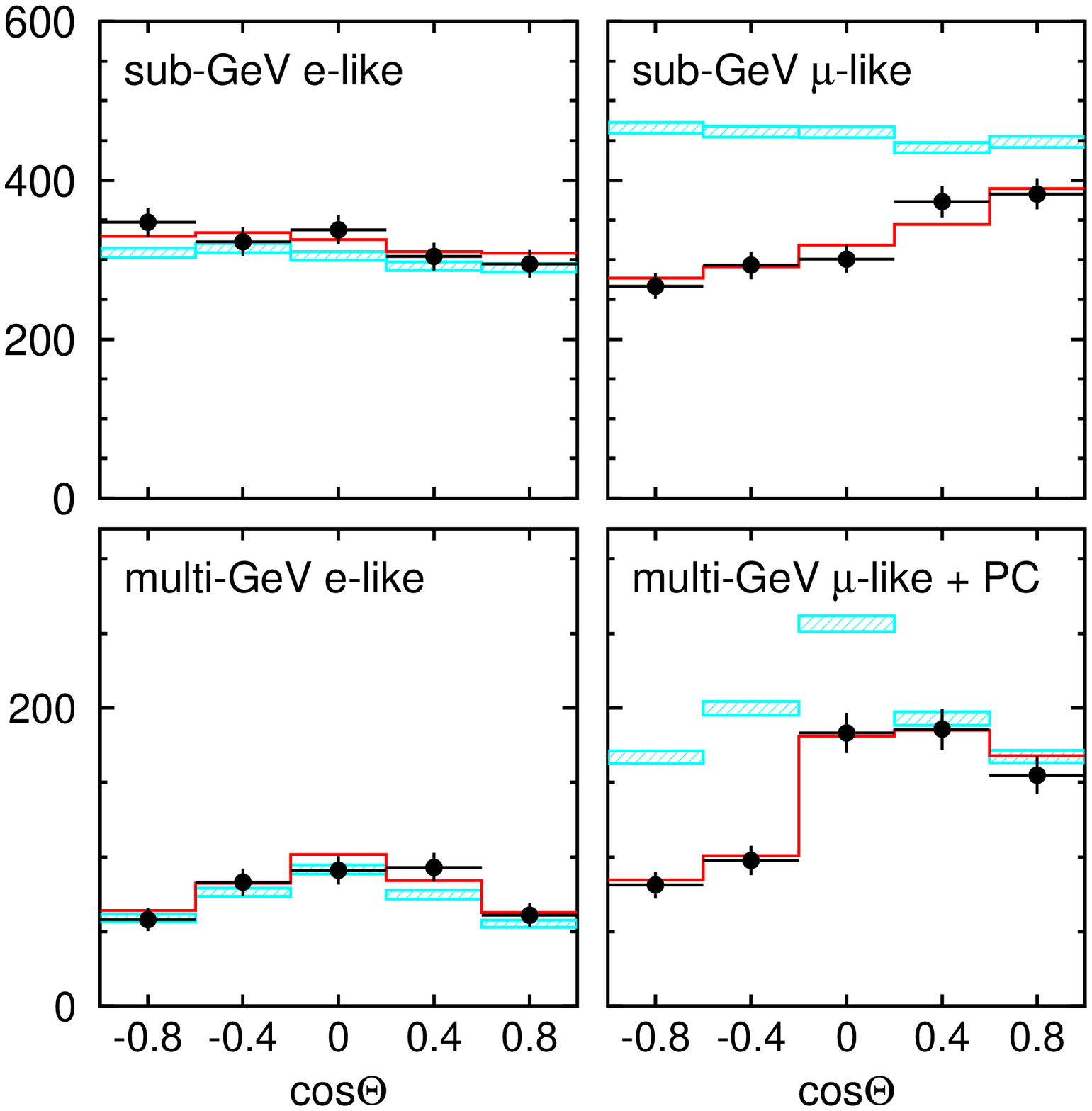,bb= 0 0 550 550,width=9.0cm}}
  \vspace*{13pt}
    \fcaption{Angular distributions for $e$-like (left) and $\mu$-like
(right) events, for sub-GeV (top) and multi-GeV (bottom) samples.  The
bars show the MC no-oscillation prediction with statistical errors,
and the line shows the oscillation prediction for the best-fit
parameters, $\sin^2 2\theta = 1.0$ and $\Delta m^2 = 3.5\times10^{-3}$
eV$^2$.  } \label{fig:angdist} \end{figure}

To study the pathlength dependence of the disappearance of atmospheric
muon neutrinos, we can study the flux as a function of zenith
angle: the Cherenkov ring direction is correlated with the neutrino
direction.  Angular distributions for FC and PC events are shown in
Figure~\ref{fig:angdist}; $\cos\theta=-1$ corresponds to upward-going
neutrinos and $\cos\theta=+1$ corresponds to downward-going.  The
oscillation prediction for the best fit parameters (see the next
section) for $\nu_{\mu}\rightarrow\nu_{\tau}$ oscillation is
superimposed.

\begin{figure}
    \centering
\mbox{\epsfig{figure=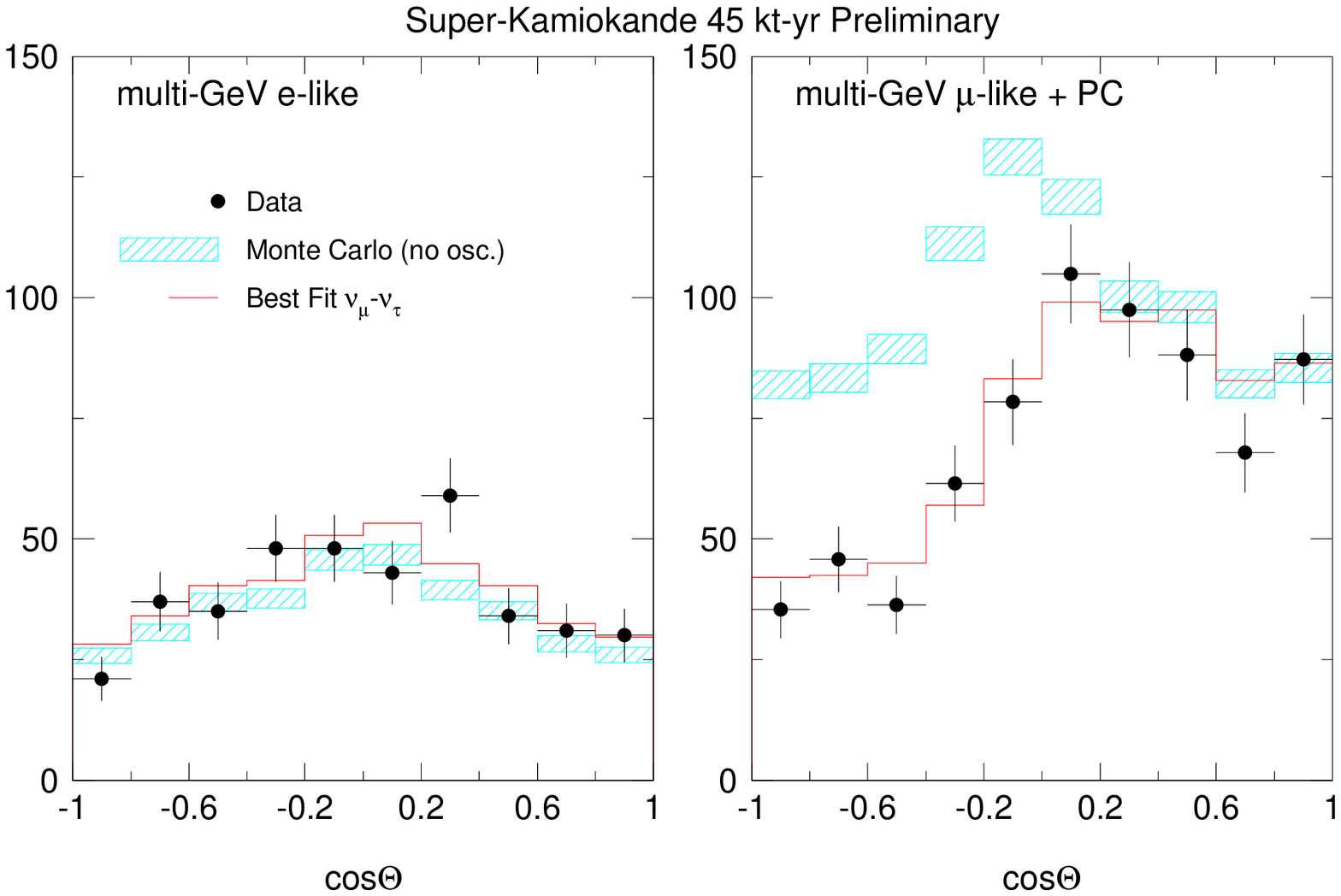,bb= 0 240 550 650,width=11.0cm}}
    \fcaption{Angular distributions for $e$-like (left) and $\mu$-like
(right) multi-GeV events (FC+PC), divided into 10 angular bins.}
\label{fig:10bin} \end{figure}

For multi-GeV events, Figure~\ref{fig:10bin} shows the angular
distributions in 10 bins, showing that the oscillation prediction matches
the data through the rapidly changing region near the horizon.

The zenith angle asymmetry $A$, defined as $A~=~\frac{U-D}{U+D}$, is
shown as a function of momentum for $e$-like and $\mu$-like events.
The value of the asymmetry for multi-GeV events (including PC) is
$A_{data}= -0.31~\pm~0.04$, more than 7$\sigma$ from the expected value
of zero, $A_{MC}= 0.01~\pm~0.01$.

  \begin{figure}
    \centering
   \mbox{\epsfig{figure=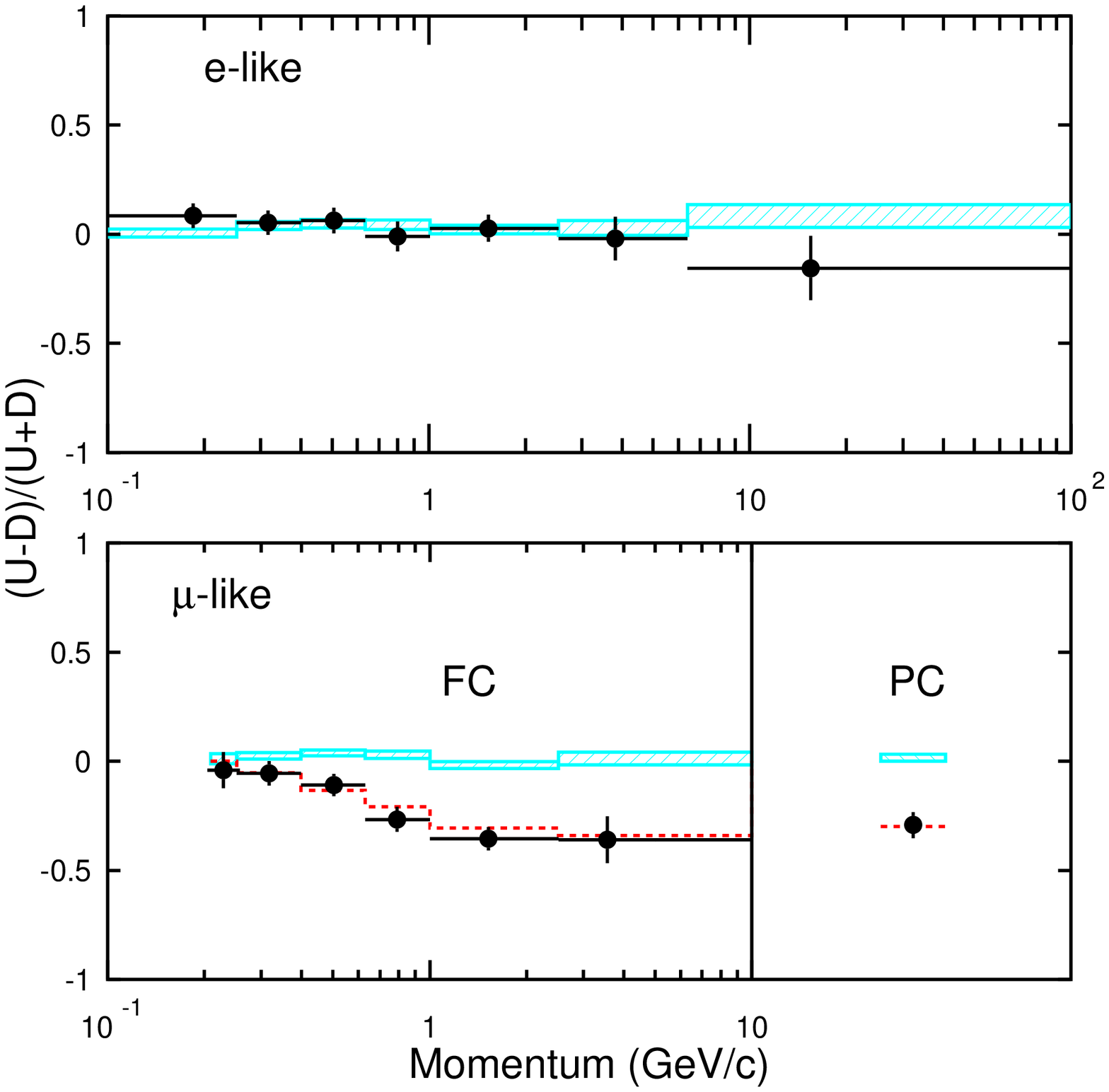,width=9.0cm}}
    \fcaption{Asymmetry as a function of momentum for $e$-like and
      $\mu$-like events.  For PC events the average neutrino energy is
      15~GeV (the produced lepton exits and does not deposit all of
      its energy. The dashed line shows the oscillation prediction for
      the best-fit parameters.}
\label{fig:asym}
\end{figure}

  \subsubsection{Oscillation Fit}\label{chi2}

The oscillation fit is performed by constructing a $\chi^2$ quantity,

\begin{equation} \label{eqn:chi2def}
\chi^2=\sum_{p,\cos\theta,e,\mu}\frac{(N_{data}-N_{osc})^2}{\sigma^2}+\sum_j \frac{\epsilon_j^2}{\sigma_j^2}.
\end{equation}
The sum is over five bins in $\cos\theta$, seven bins in $\log p$, where $p$ is
momentum, and two particle ID bins ($e$ and $\mu$).  $N_{data}$ is the
number of data events in the bin, 
$\sigma$ is the statistical error on data and MC,
and $N_{osc}$ is a weighted sum
over events:
\begin{equation}
N_{osc}=\frac{{\mathcal{L}}_{data}}{{\mathcal{L}}_{MC}}\sum_{\textrm{events}}
w(\sin^22\theta,\Delta m^2,\alpha,\epsilon_j),
\end{equation}
where $\alpha$ is the flux normalization, $\sin^22\theta$ and $\Delta
m^2$ are the oscillation parameters, and ${\mathcal L}_{data}$ and
${\mathcal L}_{MC}$ are the data and MC livetimes.  The $\epsilon_j$
values and their corresponding $\sigma_j$'s take into account 
systematic uncertainties.  Table~\ref{tab:epsilons} lists these systematic
parameters, their estimated sigmas and their best-fit values.

  \begin{table}[h]
    \centering
   \vspace*{13pt}
  \tcaption{Estimated systematic parameters.  Note that the
flux normalization $\alpha$ is not constrained in the fit.}
 \label{tab:epsilons}
  \small
  \begin{tabular}{||c|c|c||}\hline\hline
 Systematic parameter & Estimated $\sigma$ & Value at best-fit\\ \hline
 Flux normalization, $\alpha$ & ($\sim$ 25 \%)& 8.4\%\\
 Energy spectrum power law, $E^{2.7\pm \delta}$ & 0.05 & -0.004\\
 $\mu$-like/$e$-like ratio, sub-GeV, $\beta_s$& 8\% & 1.5\%\\
 $\mu$-like/$e$-like ratio, multi-GeV, $\beta_m$& 12\% & -12.5\%\\
 FC/PC ratio, $\rho$ & 8\% & -2.9\% \\
 Zenith angle asymmetry, sub-GeV, $\eta_s$ & 2.4\%  & -0.4\%\\
 Zenith angle asymmetry, multi-GeV, $\eta_m$ & 2.7\% & -0.2\%\\
 $<L/E_{\nu}>$,  $\lambda$ &15\% & -1.1\%\\
  \hline \hline
  \end{tabular} 
  \end{table}

To build an allowed region:  

\begin{itemize}
\item This
$\chi^2$ quantity is minimized at each $\sin^2 2\theta$, $\Delta m^2$
value, with respect to the systematic parameters $\epsilon_j$.
\item The minimum $\chi^2$ in $\sin^2 2\theta$, $\Delta m^2$
parameter space, $\chi^2_{min}$ is found.
\item A confidence level map is built as a function of
$\chi^2-\chi^2_{min}$.  The unphysical region is taken into account
according to a two-dimensional generalization of the Particle Data
Book prescription~\cite{pdb,mmthesis}.
\end{itemize}

\begin{figure}
    \centering
    \mbox{\epsfig{figure=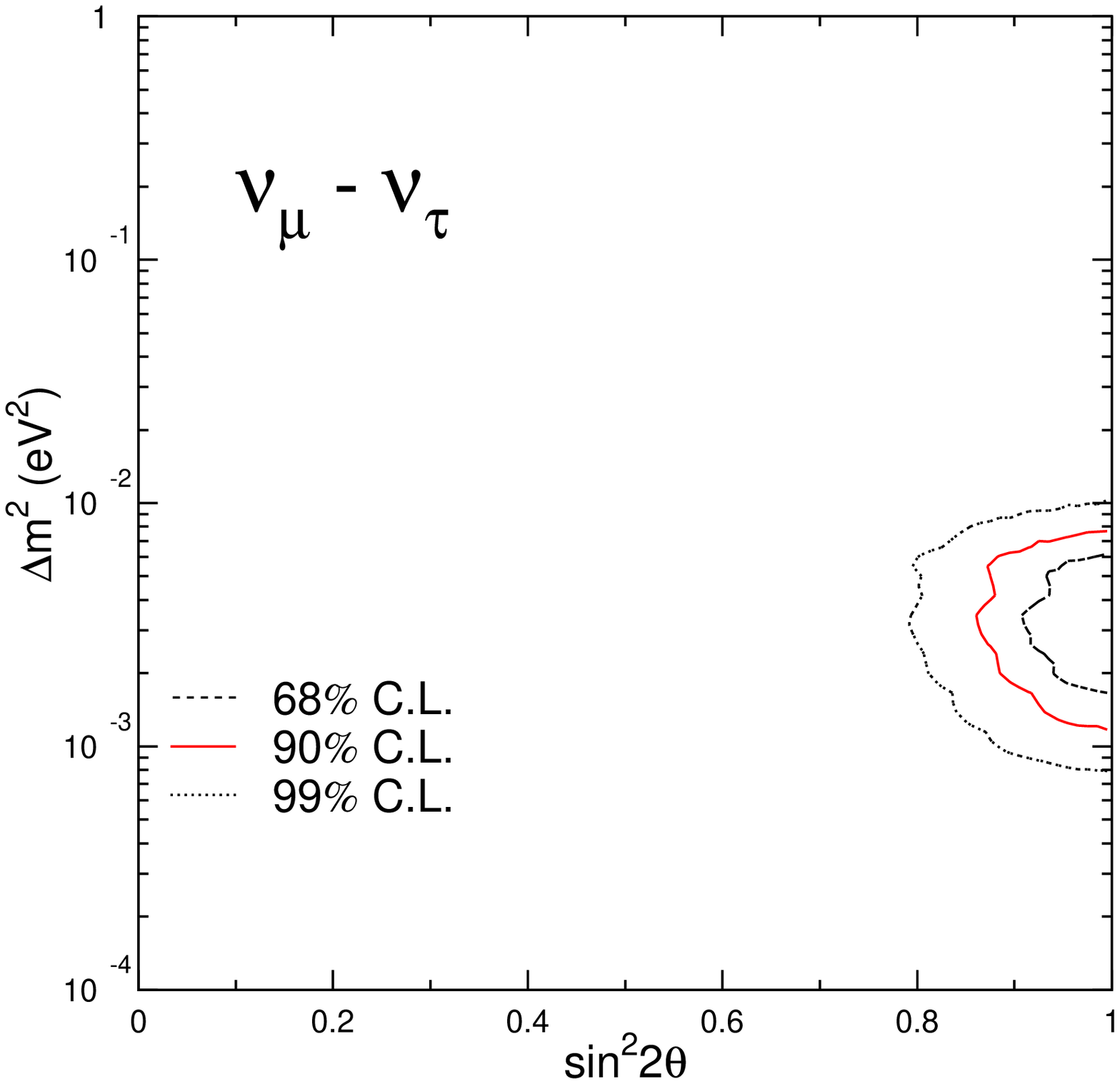,bb= 0 130 550 650,width=9.0cm}}
    \fcaption{Allowed regions for $\nu_{\mu}\rightarrow\nu_{\tau}$ oscillations, for FC and PC data.}
    \label{fig:mutau}
  \end{figure}

\begin{figure}
    \centering
    \mbox{\epsfig{figure=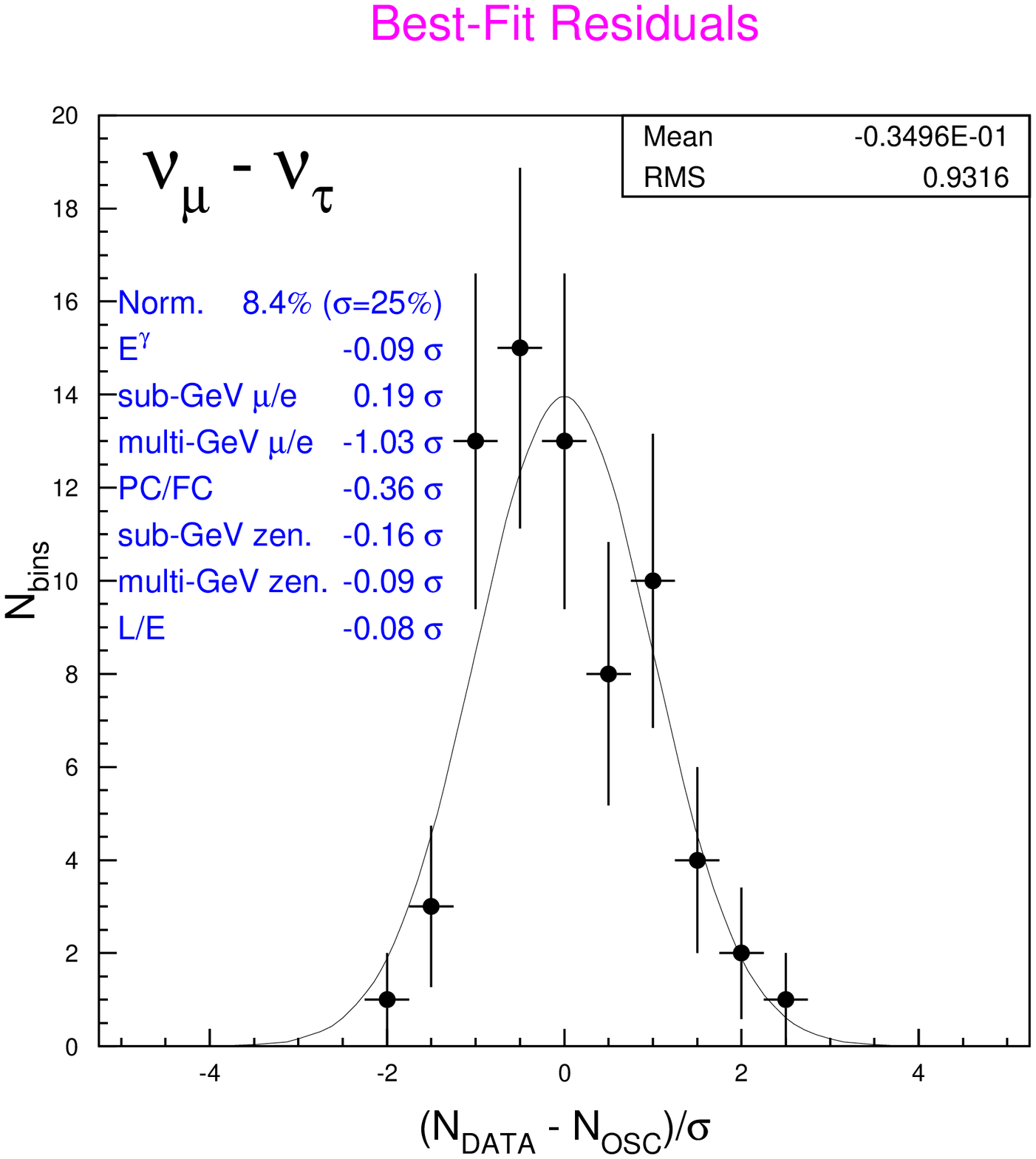,bb= 0 110 550 750,width=9.0cm}}
    \fcaption{Fit residuals: distribution of $\chi^2$ residuals for
the 70 bins in the equation~\ref{eqn:chi2def} sum at $\chi^2_{min}$.
The table on the left indicates the $\epsilon_j$ best-fit values.}
\label{fig:resid} \end{figure}

Figure~\ref{fig:mutau} shows the allowed regions for 68\%, 90\% and
99\% C.L. for $\nu_{\mu}\rightarrow\nu_{\tau}$ oscillation.  
For this case, 90\% C.L. corresponds to $\chi^2_{min}+5.3$.  The
best-fit oscillation parameter values in the physical region are
$\sin^2 2\theta=1.0$ and $\Delta m^2 = 3.5\times 10^{-3}$ eV$^2$,
corresponding to $\chi^2/d.o.f. = 62/67$.
Figure~\ref{fig:resid} shows the distribution of residuals for the 70
terms in the $\chi^2$ sum at $\chi^2_{min}$; it is well fit by a normalized
Gaussian.  All of the best-fit $\epsilon_j$ values lie within their
estimated uncertainties at the 1~$\sigma$ level.  The $\chi^2/d.o.f.$ value
for the no-oscillation hypothesis is 175/69, corresponding to a
probability of less than 0.0001\%.

\begin{figure}
    \centering
    \mbox{\epsfig{figure=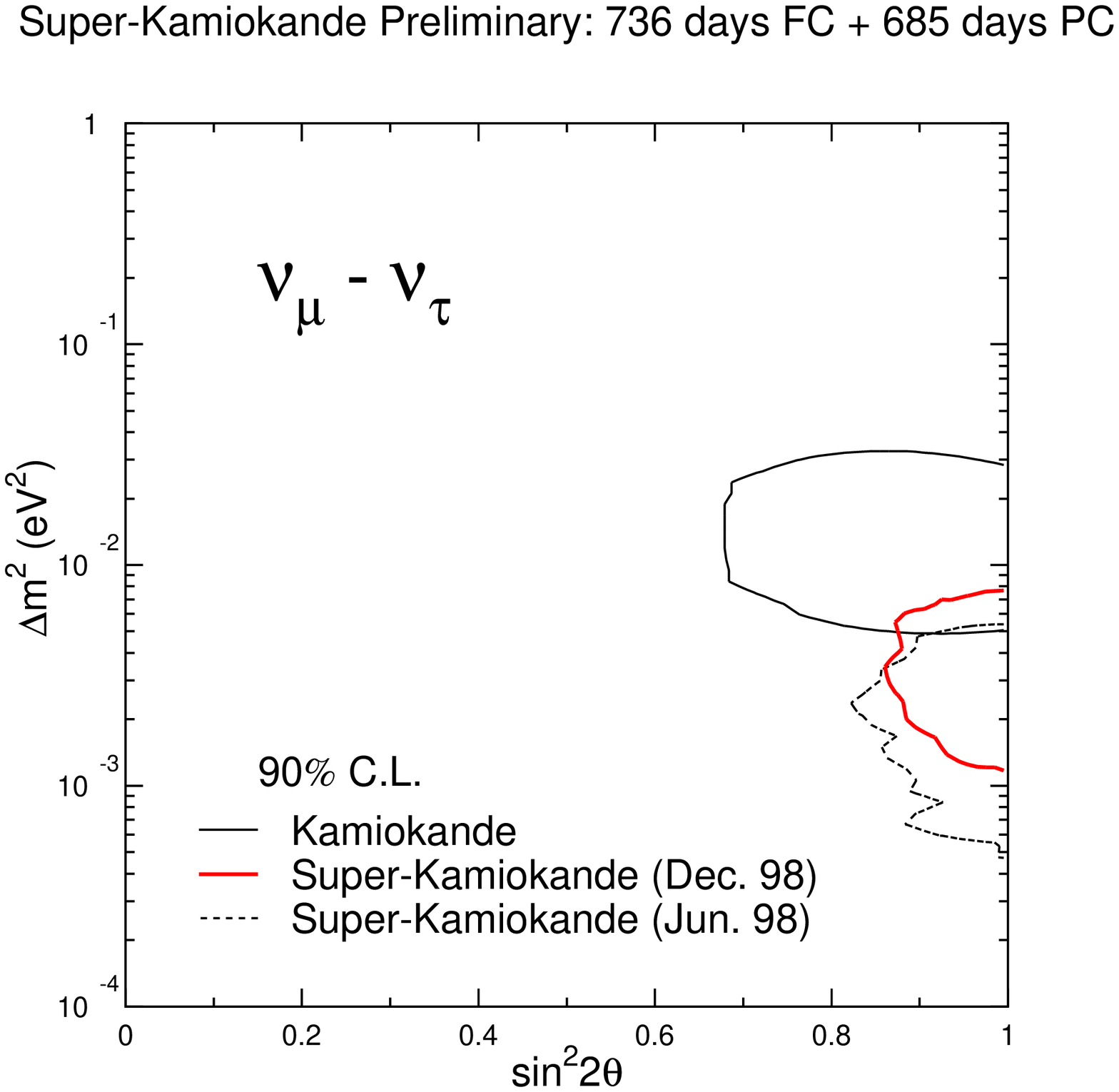,bb= 0 130 550 650,width=9.0cm}}
    \fcaption{90 \% C.L. allowed regions compared with the June 1998 Super-K
result and the Kamiokande result.}
    \label{fig:overlay}
  \end{figure}

Figure~\ref{fig:overlay} shows the new allowed region compared with
the June 1998 region and the old Kamiokande result\cite{kam}.

  \subsection{Upward-going Muons}
  
  We can get neutrino oscillation information from the upward-going
  muon (upmu) sample by comparing the angular distribution to the
  expected one.  In addition, although we can get no direct energy
  information, we do know that the stopping and through-going samples
  come from different parent neutrino energy distributions, so that
  the stopping to through-going ratio is a way of testing the energy
  dependence of the oscillation hypothesis.

\begin{figure}
\centering \begin{minipage}[t]{7.5cm} \vspace*{13pt}
  \epsfig{figure=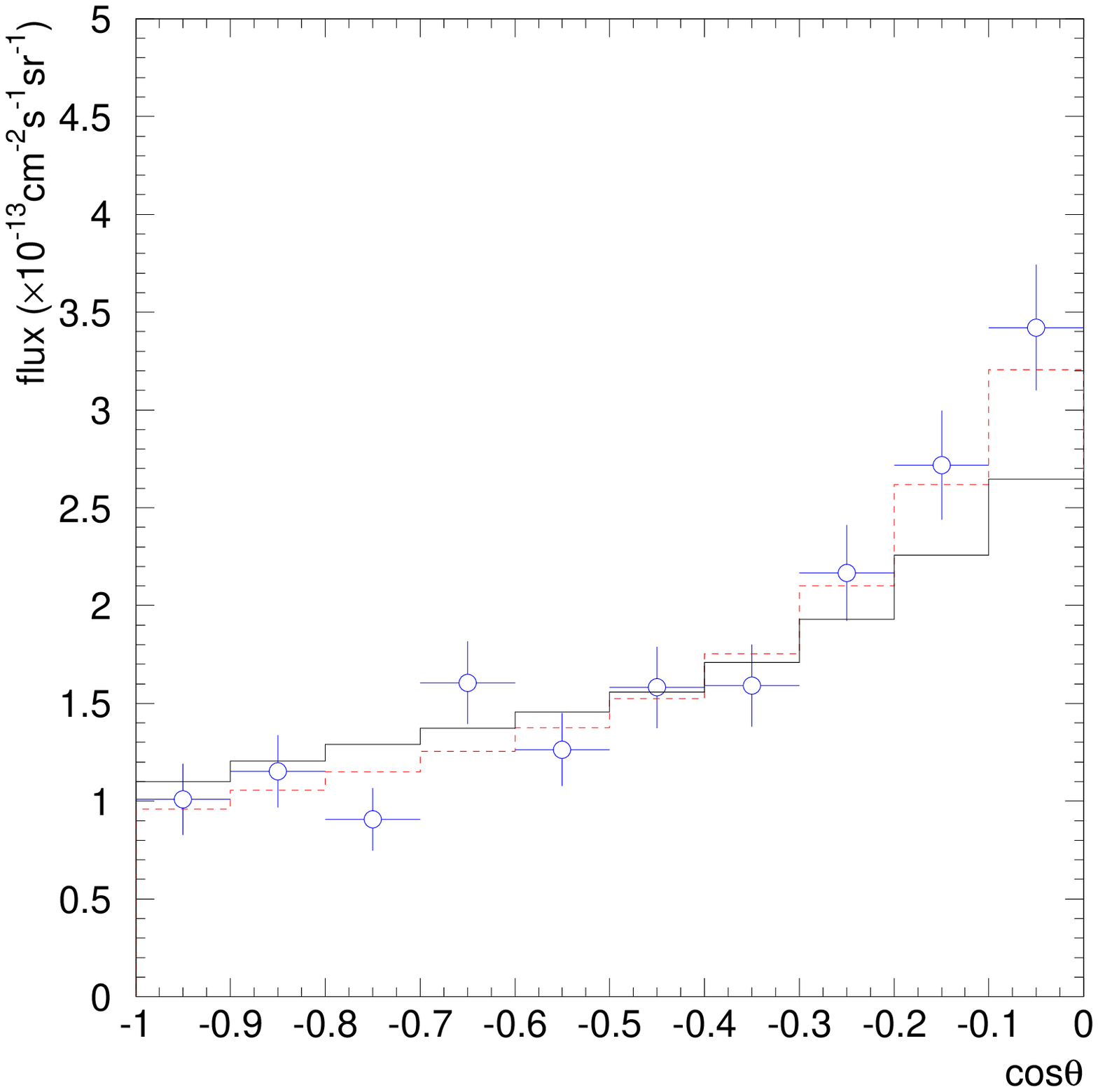,bb= 5 150 551 652,width=7.5cm}
  \vspace*{13pt} \end{minipage} \begin{minipage}[t]{7.5cm}
  \vspace*{13pt} \epsfig{figure=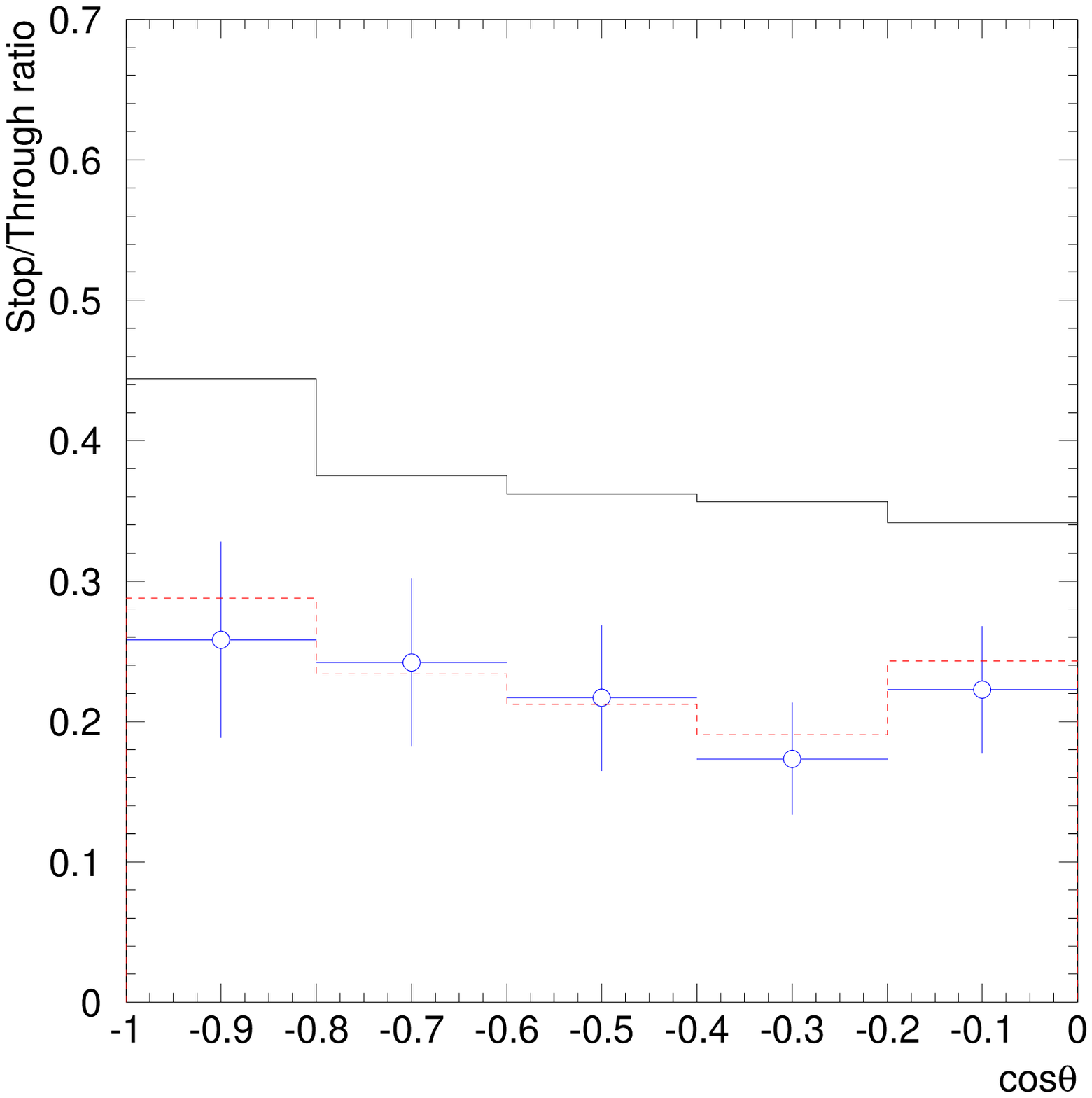,bb= 5 150 551
  652,width=7.5cm} \vspace*{13pt} \end{minipage} \vspace*{13pt}
  \fcaption{Left: Angular distribution of through-going upward muons.
  The circles represent the data, the solid line represents the
  normalized no-oscillation flux prediction, and the dashed line
  represents the best fit prediction for oscillations.  Right: ratio
  of stopping to through-going muons as a function of $\cos\theta$. }
  \label{fig:upgoers}
\end{figure}

  The through-going upward muon angular distribution is shown in 
  Figure~\ref{fig:upgoers}.  The overall stop/through ratio is given by
$\frac{N_{stop}}{N_{thru}}= 0.22 \pm 0.03$,
compared to the expected value (using the Honda flux\cite{honda}) of 
$\frac{N_{stop}}{N_{thru}}= 0.37 \pm 0.04$ for the no-oscillation case.
The stop/through ratio as a function of angle is shown in
Figure~\ref{fig:upgoers}.

\begin{figure}
    \centering
    \mbox{\epsfig{figure=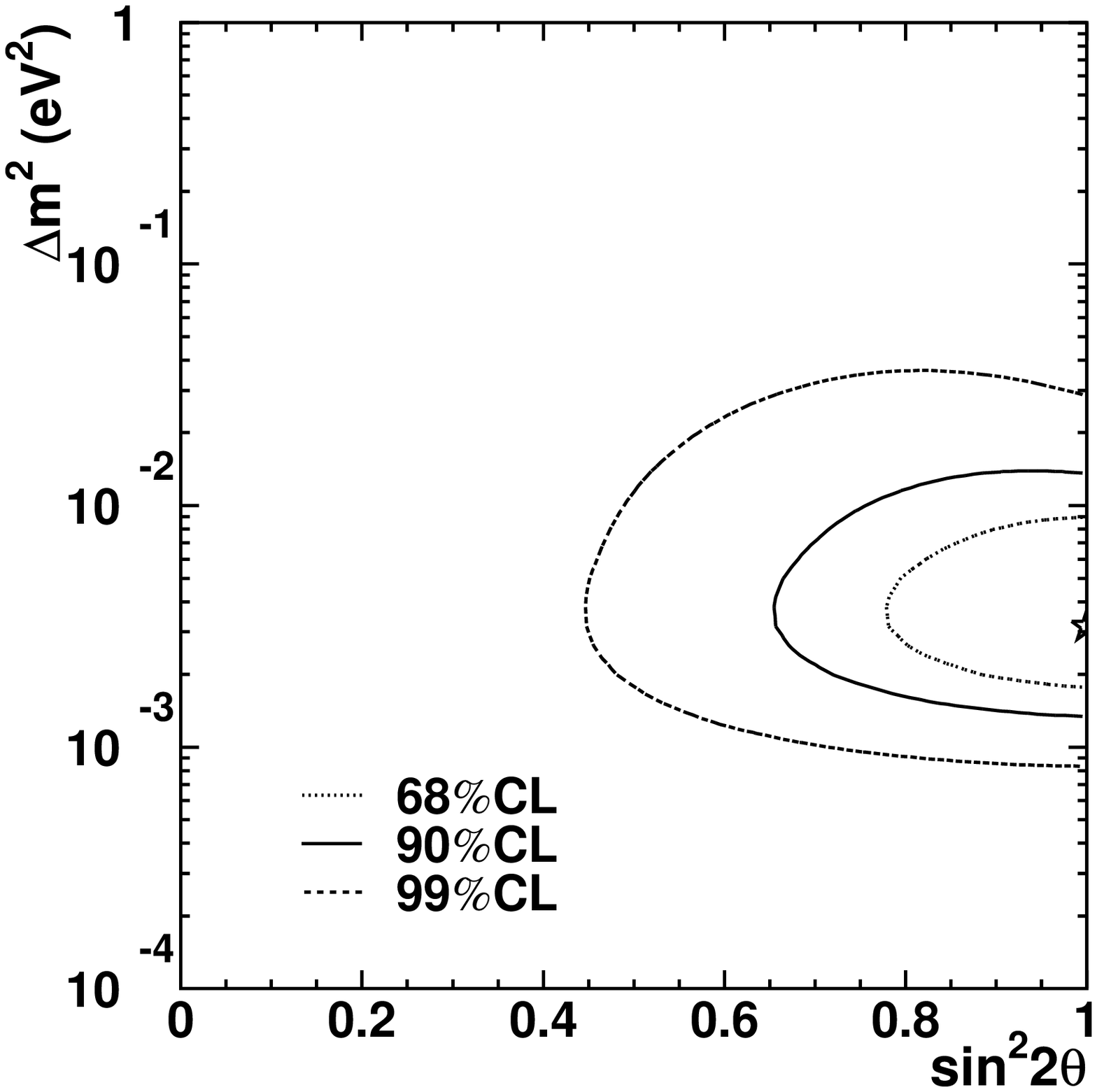,bb= 5 150 551 652,width=9.0cm}}

    \fcaption{$\nu_{\mu}\rightarrow \nu_{\tau}$ allowed regions for
    upward-going muons.}  \label{fig:upmu_allowed} \end{figure}

The oscillation fit is done by construction of a $\chi^2$ similar
to equation~\ref{eqn:chi2def}:

\begin{equation} \label{eqn:chi2def2}
\chi^2=\sum_{i=1}^{15}\frac{(\Phi_{data}-\Phi_{osc})^2}{\sigma^2}+\left(\frac{\alpha}{\sigma_{\alpha}}\right)^2 +\left(\frac{\eta}{\sigma_{\eta}}\right)^2,
\end{equation}
\noindent
where the sum is over 10 zenith angle bins for through-going muons and
five zenith angle bins for stopping muons.  For this case, the weighted
MC prediction is replaced by a flux calculation\cite{honda}, and
\begin{equation}
\Phi_{osc} = \Phi_{calc}(\sin^2 2\theta,\Delta m^2)\ast (1+\alpha) \textrm{~~for through-going muons, and}
\end{equation}

\begin{equation}
\Phi_{osc} = \Phi_{calc}(\sin^2 2\theta,\Delta m^2)\ast (1+\alpha)(1+\eta) \textrm{~~for stopping muons.}
\end{equation}
\noindent
$\alpha$ is the absolute flux normalization ($\sigma_{\alpha}\sim
22\%$) and $\eta$ corresponds to the relative stopping and
through-going uncertainty ($\sigma_{\eta}\sim 13\%$).  The 90\%
C.L. allowed region (including both through-going and stopping upmu
information) is shown in Figure~\ref{fig:upmu_allowed}.  The best-fit
oscillation parameters are $\sin^2 2\theta=1$ and $\Delta m^2 = 3.2
\times 10^{-3}$~eV$^2$, corresponding to $\chi^2/d.o.f.=8/13$.

  \subsection{Combined Analysis}

We can combine FC/PC and upmu data\cite{upmudpf} for a
global fit by constructing a $\chi^2$ quantity including information from
both the FC/PC and upmu $\chi^2$ quantities
(equations~\ref{eqn:chi2def} and~\ref{eqn:chi2def2}):

\begin{equation} \label{eqn:chi2def3}
\chi^2=\sum\frac{(N_{data}-N_{osc})^2}{\sigma^2}+\sum_j \frac{\epsilon_j^2}{\sigma_j^2},
\end{equation}
\noindent
where the sum is over the 70 FC/PC bins, and the 15 upmu bins.  For
FC/PC bins, $N_{osc}$ represents the weighted MC term of the FC/PC
fit, whereas for the upmu bins, $N_{osc}$ represents the flux
prediction.  The systematic parameters $\epsilon_j$ include the same
FC/PC terms as before; there is a common flux normalization parameter
$\alpha$ (unconstrained, as for the FC/PC fit) and a common spectral
index uncertainty $\delta$.  In addition the parameter $\eta$ of the
upmu fit is replaced by two parameters $\eta_1$ and $\eta_2$, each
with an assigned $\sigma$ of 7\%, representing the relative stopping
and through-going uncertainty and the relative FC/PC and upmu
uncertainty respectively.

\begin{figure}
    \centering
  \vspace*{13pt}
    \mbox{\epsfig{figure=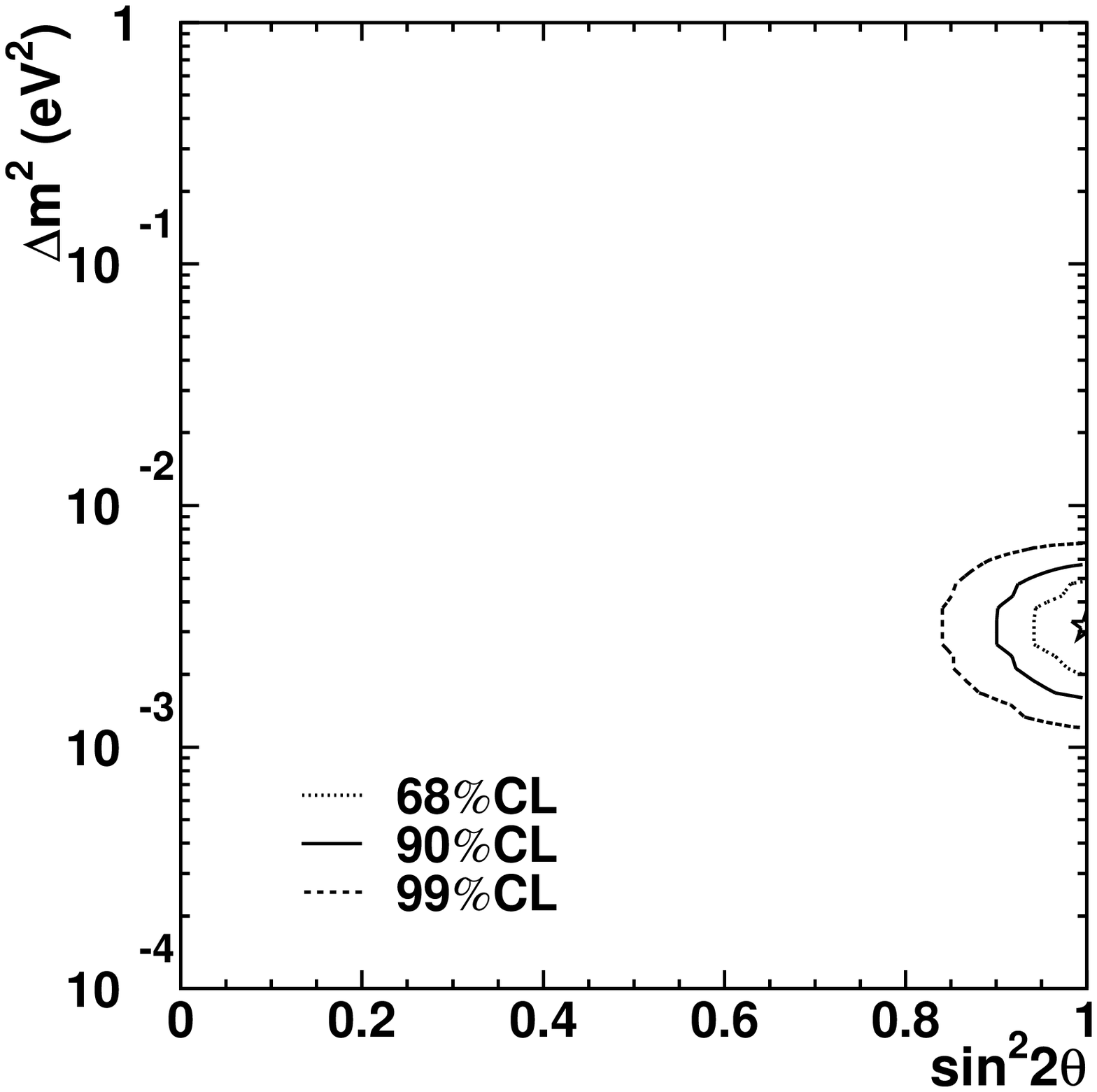,bb= 5 130 551 652,width=9.0cm}}
    \fcaption{Allowed regions using combined information from FC, PC and upward-going
muons.}
    \label{fig:combined}
  \end{figure}

The resulting $\nu_{\mu}\rightarrow\nu_{\tau}$ confidence level
contours are shown in Figure~\ref{fig:combined}.  The best-fit value
is at $\sin^2 2\theta=1$ and $\Delta m^2=3.2\times 10^{-3}$ eV$^2$,
corresponding to $\chi^2/d.o.f.$ of 70.2/82.

\subsection{FC Multiple Rings}

An additional FC sample can be used to check for consistency of
oscillation behavior: the multiple ring component of the FC sample
(all FC results so far have been for single rings, to tag the simplest
quasielastic interactions).  To produce
enriched muon and electron flavor 2-ring samples,
sub-GeV 2-ring events were required to satisfy the following
conditions: 

\begin{itemize}
\item $p>300$ MeV.
\item For $\nu_e$-enriched: 2 $e$-like rings not in the $\pi^0$
peak (see Section~4.2.2) or 1~$e$-like ring and 1~$\mu$-like ring with
the $e$-like ring having higher momentum.  For either case, there must
no muon decay electrons.
\item For $\nu_{\mu}$-enriched: 2~$\mu$-like rings or 1~$e$-like ring
and 1~$\mu$-like ring with the $\mu$-like ring having higher momentum.
For either case, there must be at least one muon decay electron.
\end{itemize}

\begin{figure}
\centering
\begin{minipage}[t]{7cm} 
    \epsfig{figure=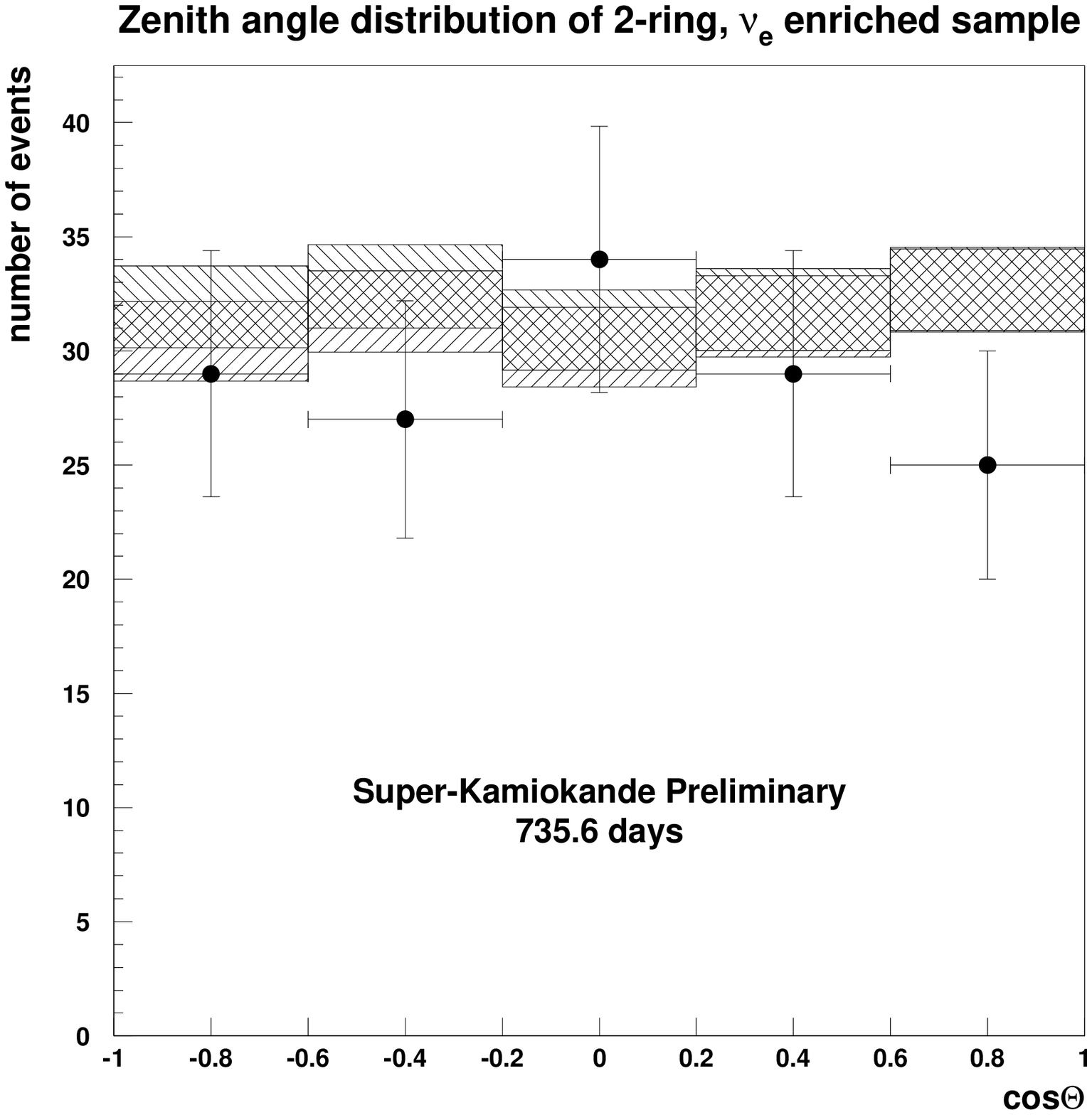,bb= 5 150 551 652,width=7.0cm}
\end{minipage}
\begin{minipage}[t]{7cm}
    \epsfig{figure=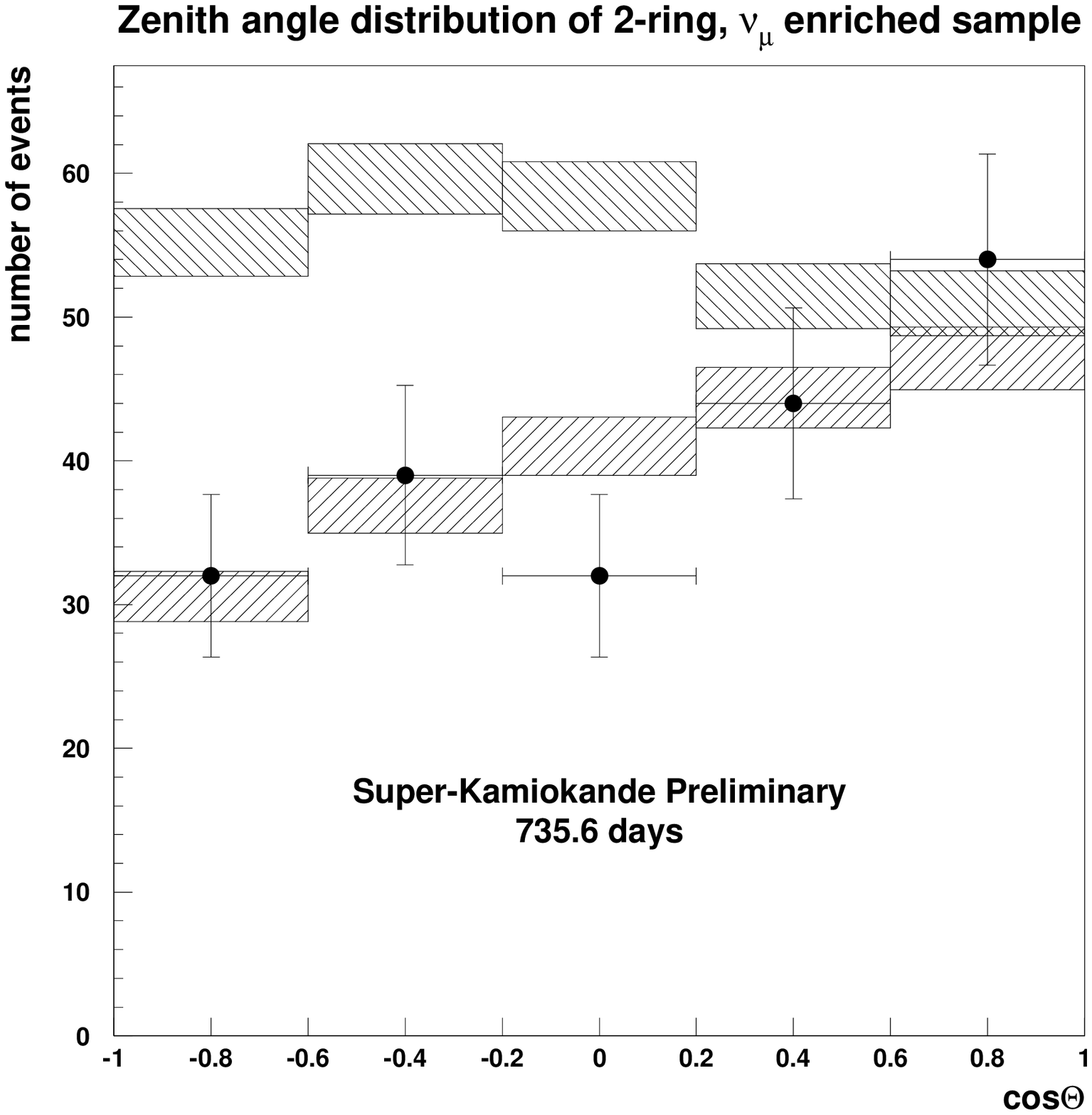,bb= 5 150 551 652,width=7.0cm}
\end{minipage}
\vspace*{13pt}
    \fcaption{Left: angular distribution for $\nu_e$ enriched sample of
multiple-ring events. The left-hatched bars show the MC prediction for
no oscillation and the right-hatched bars show the prediction
for the oscillation parameters $\sin^2 2\theta=1$ and $\Delta m^2=3.5\times 10^{-3}$~eV$^2$.  Right:  $\nu_{\mu}$-enriched sample.}  \label{fig:mring}
  \end{figure}

According to MC, the enriched $\nu_e$ sample is 60\% $\nu_e$ CC, 8\%
$\nu_{\mu}$ CC and 32\% NC, and the enriched $\nu_{\mu}$ sample is 89\%
$\nu_{\mu}$ CC, 2\% $\nu_{e}$ CC and 8\% NC. The angular distributions
for this preliminary analysis are shown in Figure~\ref{fig:mring}.
Clearly, the results are consistent with the oscillation hypothesis.

  \section{Oscillation Interpretation} 

For all of the allowed regions so far, we have been assuming that the
muon disappearance results from $\nu_{\mu}$ to $\nu_{\tau}$
oscillation.  However, this assumption is not necessarily true.  What
can we say about the flavors involved in the oscillation?

  \subsection{$\nu_{\mu} \rightarrow \nu_{e}$}

\begin{figure}
    \centering
    \mbox{\epsfig{figure=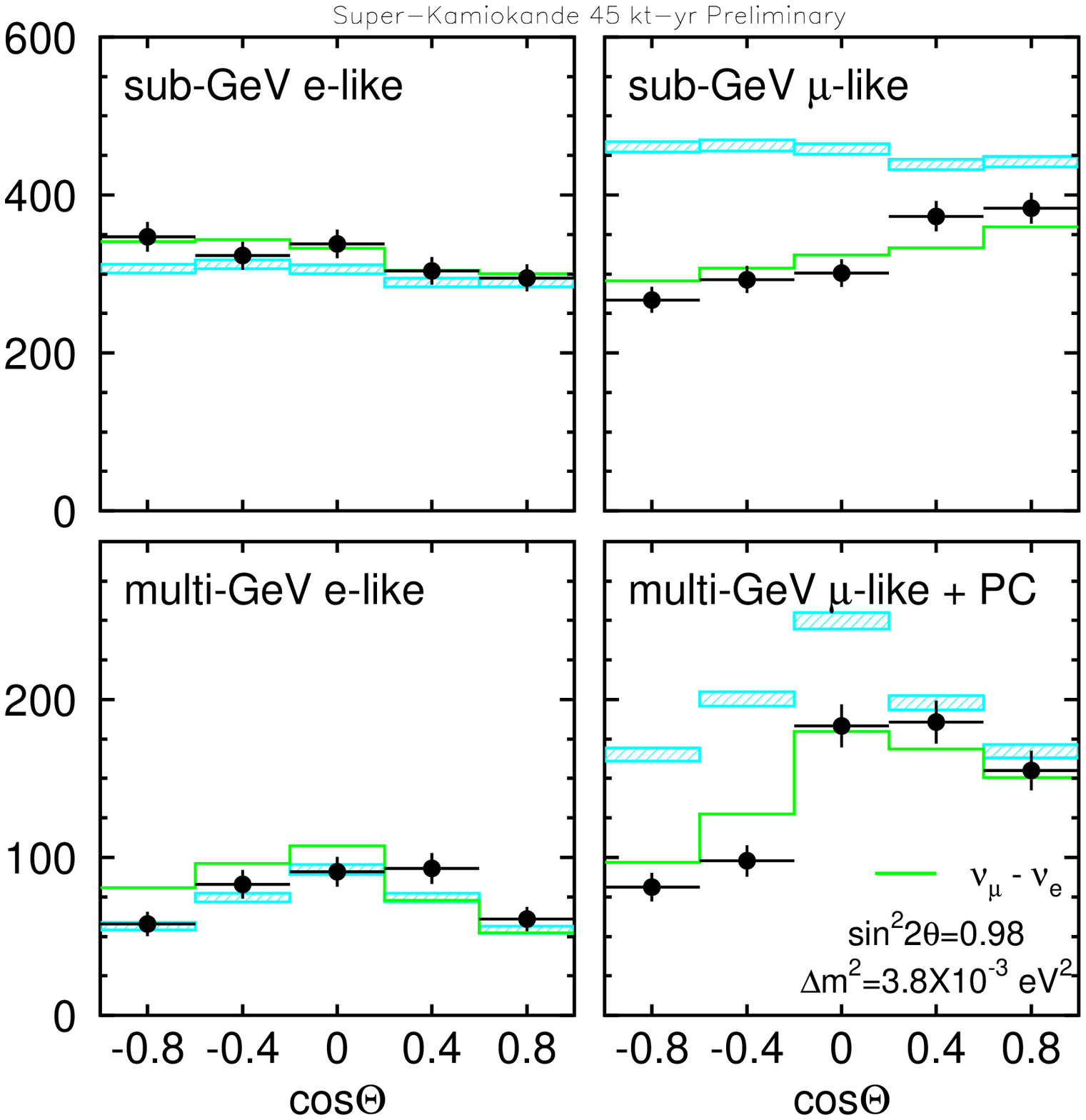,bb= 0 130 550 650,width=9.0cm}}
    \fcaption{Angular distributions with the best fit for 
$\nu_{\mu}\rightarrow \nu_{e}$ superimposed.}
    \label{fig:mue}
  \end{figure}

Are the muon neutrinos oscillating into electron neutrinos?  This is a
fairly easy question to answer -- the Super-K angular distribution by
itself is inconsistent with pure $\nu_{\mu} \rightarrow \nu_{e}$
oscillation.  This assumption (including matter effects in the Earth)
yields a very poor fit: $\chi^2/d.o.f.=110/67$ (see
Figure~\ref{fig:mue}).  In addition, the CHOOZ experiment in France
rules out this hypothesis (for $\bar{\nu_e}$ disappearance)\cite{chooz}.

  \subsection{$\nu_{\mu} \rightarrow \nu_{s}$}
  
  Could the muon neutrinos be oscillating into a non-weakly
  interacting sterile neutrino?  With Super-K data alone, it is hard to
  tell this hypothesis apart from the $\nu_{\mu} \rightarrow
  \nu_{\tau}$ hypothesis; we expect only tens of CC $\nu_{\tau}$
  interactions in the current sample, and the products of such
  interactions in our detector are nearly indistinguishable from other
  atmospheric neutrino events.  However, an oscillation to a sterile
  neutrino could nevertheless leave subtle imprints on some properties
  of our data set.  Here two preliminary studies are described.

  \subsubsection{Angular Distribution}

Due to the fact that sterile neutrinos do not exchange $Z$ bosons with
nucleons in the earth, the angular distribution for $\nu_{\mu}
\rightarrow \nu_{s}$ will be distorted with respect to the $\nu_{\mu}
\rightarrow \nu_{\tau}$ oscillation case (see, for instance, Lipari
and Lusignoli~\cite{sterile}).  This effect is expected to be largest
for high $E$ and low $\Delta m^2$ cases. The effect is, however,
smeared over a distribution of neutrino energies.

\begin{figure}
    \centering
    \mbox{\epsfig{figure=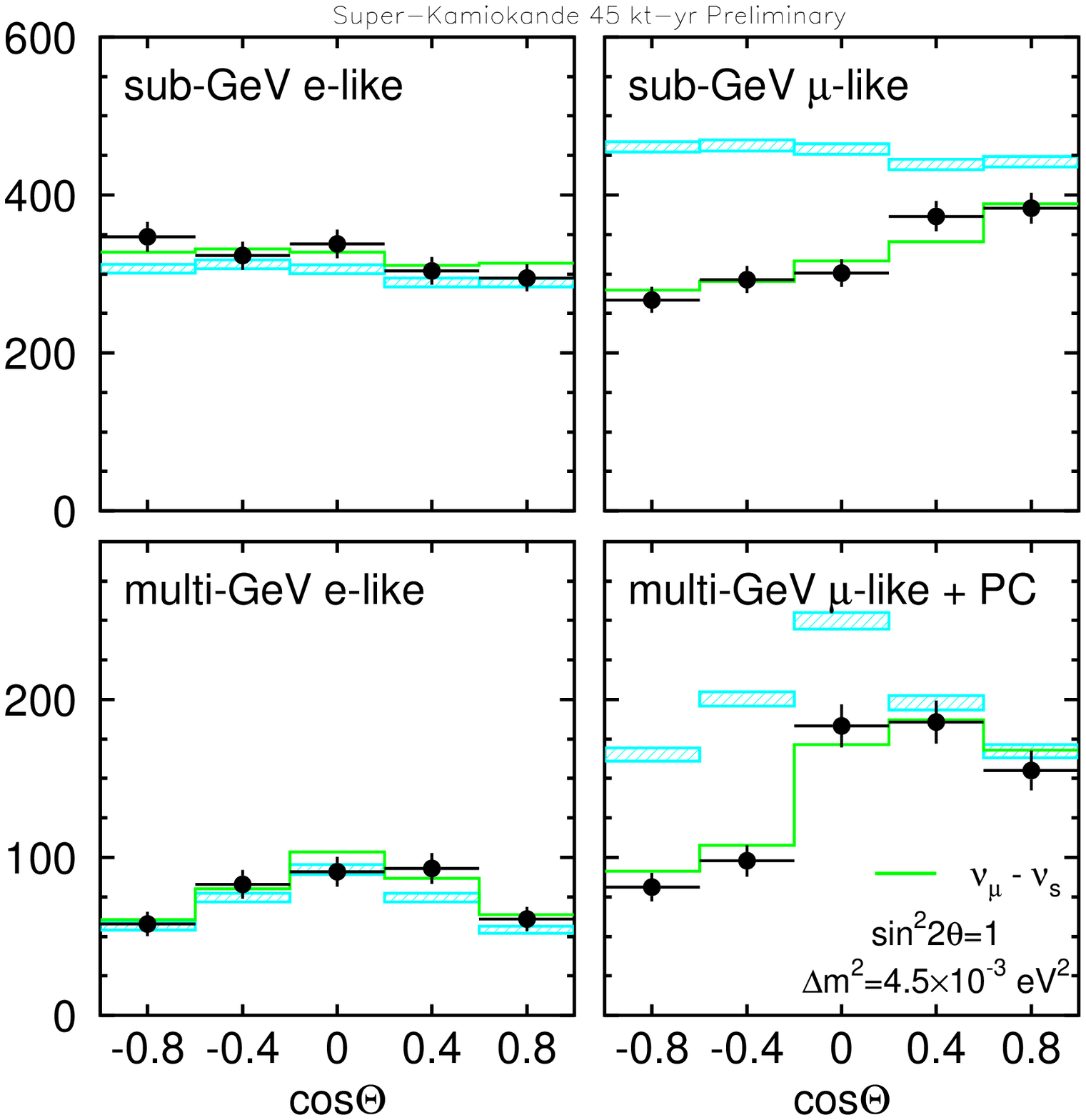,bb= 0 130 550 650,width=9.0cm}}
    \fcaption{Angular distributions with the best fit for \
$\nu_{\mu}\rightarrow \nu_{s}$ superimposed.}
    \label{fig:mus_angdist}
  \end{figure}

\begin{figure}
    \centering
    \mbox{\epsfig{figure=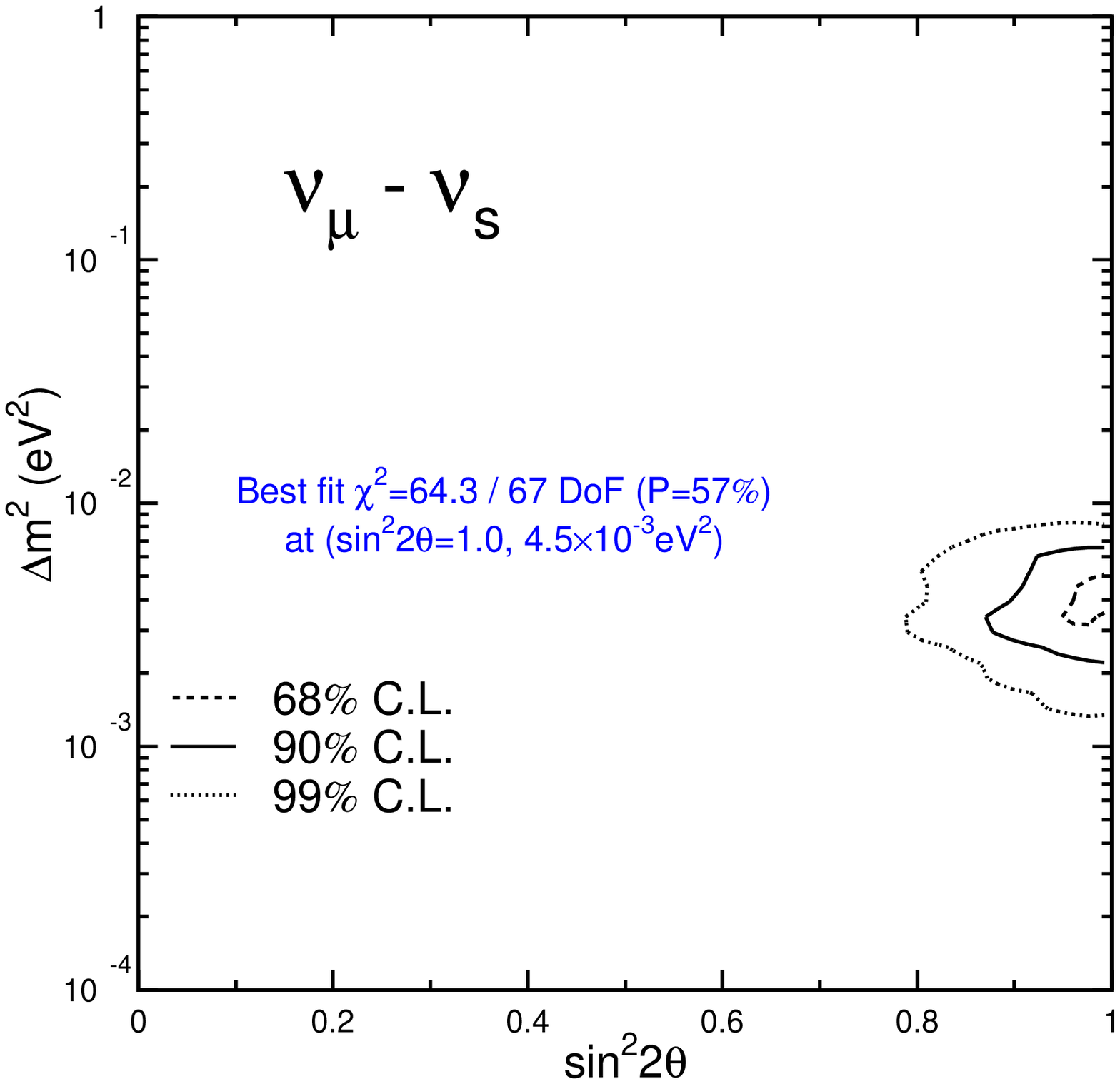,bb= 0 130 550 650,width=9.0cm}}
    \fcaption{Allowed region for \
$\nu_{\mu}\rightarrow \nu_{s}$.}
    \label{fig:mus}
  \end{figure}

  Figure~\ref{fig:mus_angdist} shows the fit for
  $\nu_{\mu}~\rightarrow~\nu_{s}$ : a good fit ($\chi^2/d.o.f.=64/67$)
  is obtained.  The corresponding allowed region is shown in
  Figure~\ref{fig:mus}; it is slightly smaller than the corresponding
  $\nu_{\mu}~\rightarrow~\nu_{\tau}$ region, due to additional
  constraints from the matter effect in the earth.  We have not yet
  included the high energy upward muon sample in this analysis.

  \subsubsection{Single Neutral Pions}\label{spi}

A rather promising way to distinguish the tau and sterile oscillation
cases is via NC neutrino interactions in the detector which produce
single neutral pions.  
Sterile neutrinos have no NC interactions; therefore,
fewer $\pi^0$ events than expected for the $\nu_{\mu}\rightarrow
\nu_{\tau}$ case would be a signature of oscillation to a sterile
neutrino.

\begin{figure}
\centering
    \epsfig{figure=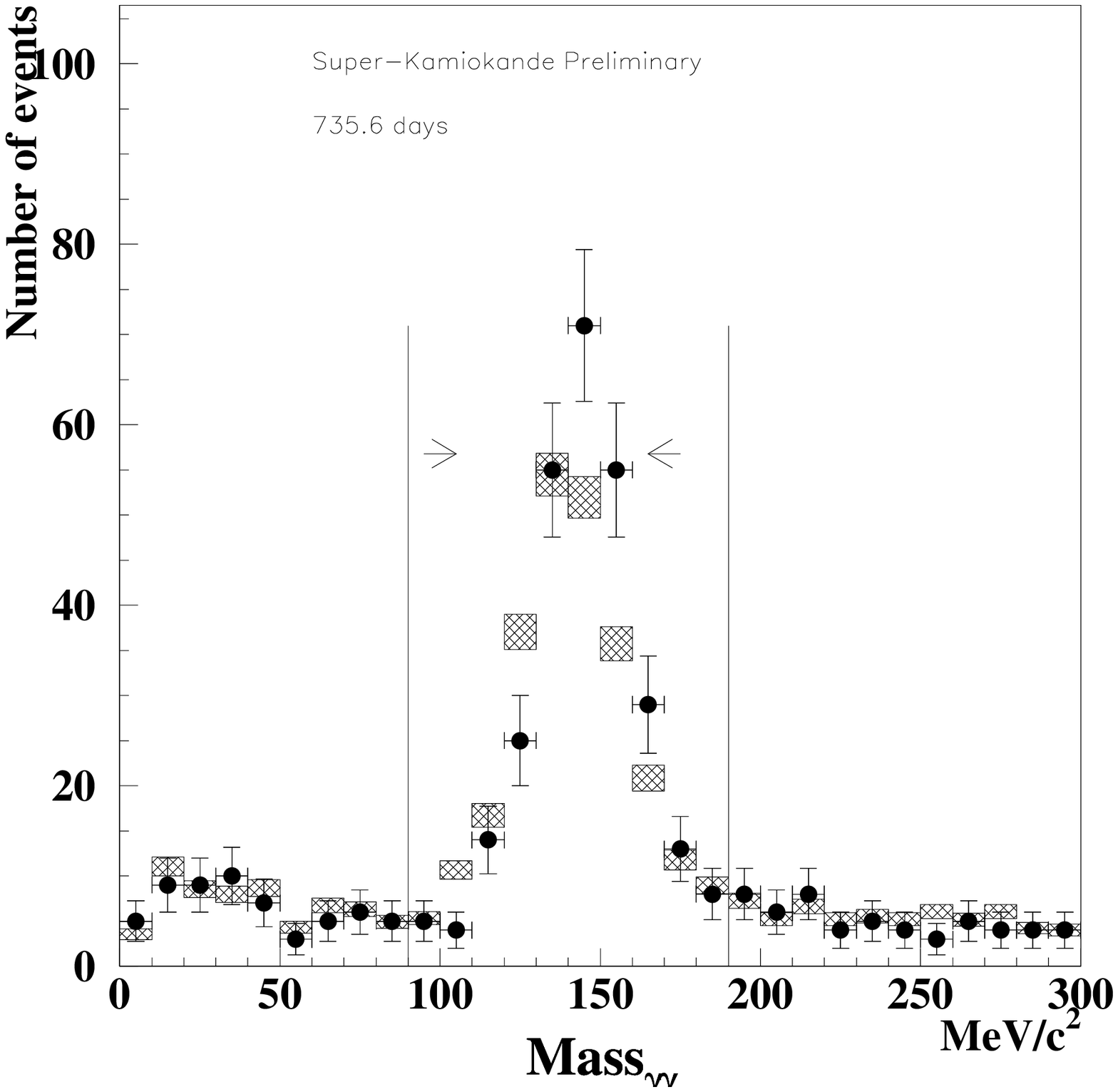,bb= 5 130 551 652,width=9.0cm}
  \vspace*{13pt}
    \fcaption{Invariant mass distribution for $\pi^0$-like events
selected according to the criteria given in the text.  The mass cut
is shown.}
    \label{fig:spi}
  \end{figure}

The signature of a $\pi^0$ NC event in the detector is a pair of
$e$-like rings ($\gamma$-induced electromagnetic showers) from the
decay of the pion, $\pi^0 \rightarrow \gamma \gamma$.  We can select a
sample of such events by selecting events with exactly two $e$-like
rings and no muon decay electrons.  Figure~\ref{fig:spi} shows the
reconstructed invariant mass distribution of this selected sample.
The $\pi^0$ peak is clearly visible.  We further reduce and enrich
the sample by making an invariant mass cut of $90~\rm{MeV}/c^2<M_{\gamma
\gamma}<190~\rm{MeV}/c^2$.  According to MC, the events surviving this cut
are an 84\% pure sample of NC interactions.  The angular distribution
of these events is shown in Figure~\ref{fig:spi}.

We can also take the ratio of $\pi^0/e$-like events and compare it to
the expected value for MC.  The FC single ring
$e$-like sample is mostly CC interactions of non-oscillated $\nu_e$'s.
This ratio of ratios should be nearly 1 if
muon neutrinos are oscillating to tau neutrinos (because tau neutrinos
interact via NC), and less than~1 if muon neutrinos are oscillating
into sterile neutrinos.  Table~\ref{tab:spi} shows the numbers.

  \begin{table}[h]
    \centering
  \vspace*{13pt}
  \tcaption{$\pi^0$ candidates.}
 \label{tab:spi}
  \small
  \begin{tabular}{||c|c|c||}\hline\hline
  &  Data & MC\\ \hline
  $\pi^0$-like & 279 & 253.6 \\
  $\pi^0$-like, background-subtracted & 231.8 & 195.9 \\
  \hline \hline
  \end{tabular} 
  \end{table}

The result is:
\begin{equation}
\frac{(\pi^0/e)_{DATA}}{(\pi^0/e)_{MC}}=1.03 \pm 0.06 (\rm{data~stat}) \pm 0.02 (\rm{MC~stat}) \pm 0.24 (\rm{sys})
\end{equation}
  
If the background is subtracted, the result is:
\begin{equation}
\frac{(\pi^0/e)_{DATA}}{(\pi^0/e)_{MC}}=1.11 \pm 0.06 (\rm{data~stat}) \pm 0.02 (\rm{MC~stat}) \pm 0.26 (\rm{sys})
\end{equation}

Notice that the systematic contribution to the uncertainty exceeds the
statistical contribution by a large factor.  The systematic
uncertainty is in fact dominated by uncertainties in the $\pi^0$
production cross-section in water.  The K2K collaboration~\cite{k2k}
hopes to reduce this uncertainty by high statistics measurements of
neutrino interactions in its near water detector.

  \section{Summary}
  
  Table~\ref{tab:summary} summarizes the status of fits of the new
  Super-K data to three possible two-flavor oscillation scenarios.  We
  obtain good fits for $\nu_{\mu}\rightarrow\nu_{\tau}$ and
  $\nu_{\mu}\rightarrow\nu_{s}$; pure two-flavor
  $\nu_{\mu}\rightarrow\nu_{e}$ is ruled out.  More work is currently
  in progress.

  \begin{table}[h] \centering \tcaption{Summary of two-flavor
    oscillation fits for new Super-K results.}\label{tab:summary}

 \vspace*{13pt}
  \small
  \begin{tabular}{||c|c|c|c||}\hline\hline
  \multicolumn{4}{||c||}{} \\ 
  \multicolumn{4}{||c||}{\textbf{No-oscillation fits}} \\ 
  \multicolumn{4}{||c||}{} \\ \hline \hline
 &   & & $\chi^2/d.o.f.$\\ \hline
FC/PC &  & & 175/69\\ 
upmu &  &  & 41/15\\
combined & &  & 214/84\\
  \hline \hline
  \multicolumn{4}{c}{} \\ \hline \hline
  \multicolumn{4}{||c||}{} \\ 
  \multicolumn{4}{||c||}{\textbf{Fits to $\nu_{\mu}\rightarrow\nu_{\tau}$}} \\ 
  \multicolumn{4}{||c||}{} \\ \hline \hline
 &  $\Delta m^2$ (eV$^2$)  & $\sin^2 2\theta$  & $\chi^2/d.o.f.$\\ \hline
FC/PC & $3.5\times 10^{-3}$ & 1.0 & 62.1/67\\ 
upmu & $3.2\times 10^{-3}$ & 1.0 & 8/13\\
combined & $3.2\times 10^{-3}$& 1.0 & 70.2/82\\
  \hline \hline
  \multicolumn{4}{c}{} \\ \hline \hline
  \multicolumn{4}{||c||}{} \\ 
  \multicolumn{4}{||c||}{\textbf{Fit to $\nu_{\mu}\rightarrow\nu_{s}$}} \\ 
  \multicolumn{4}{||c||}{} \\ \hline \hline
FC/PC & $4.5\times 10^{-3}$ & 1.0 & 64.3/67\\ 
  \hline \hline
  \multicolumn{4}{c}{} \\ \hline \hline
  \multicolumn{4}{||c||}{} \\ 
  \multicolumn{4}{||c||}{\textbf{Fit to $\nu_{\mu}\rightarrow\nu_{e}$}} \\ 
  \multicolumn{4}{||c||}{} \\ \hline \hline
FC/PC & $3.8\times 10^{-3}$ & 0.98 & 110/67\\ 
  \hline \hline

  \end{tabular} 
  \end{table}

  \section{Acknowledgements}
The Super-Kamiokande experiment is supported by the Japanese Ministry
of Education, Science, Sports and Culture and the United States
Department of Energy.

  \end{document}